\documentstyle[12pt,epsf]{article}

\newcommand{\beq}{\begin{equation}}
\newcommand{\eeq}{\end{equation}}
\newcommand{\beqa}{\begin{eqnarray}}
\newcommand{\eeqa}{\end{eqnarray}}
\newcommand{\no}{\nonumber}

\newcommand{\qq}{\qquad}
\newcommand{\mnod}{\stackrel{\circ}{m}}

\newcommand{\ba}{\begin{array}}
\newcommand{\ea}{\end{array}}
\newcommand{\tr}{\mbox{tr}}
\newcommand{\ci}{\mbox{i}}

\newcommand{\ol}{\overline}

\newcommand{\wh}{\widehat}

\newcommand{\D}{{\cal D}}

\newcommand{\dfrac}{\displaystyle \frac}

\newcommand{\nn}{\nonumber \\}

\newcommand{\del}{\partial}

\begin{document}

\hfill 

\hfill 

\bigskip\bigskip

\begin{center}

{{\Large\bf Weak hyperon decays in heavy baryon
            chiral perturbation theory: Renormalization and applications }}

\end{center}

\vspace{.4in}

\begin{center}
{\large B. Borasoy$^a$\footnote{email: borasoy@physik.tu-muenchen.de}
and G. M{\"u}ller$^b$\footnote{email: gmueller@doppler.thp.univie.ac.at }}

\bigskip

\bigskip

$^a$Physik Department\\
Technische Universit{\"a}t M{\"u}nchen\\
D-85747 Garching, Germany\\

\bigskip

$^b$Universit\"at Wien \\
     Institut f\"ur Theoretische Physik  \\
     Boltzmanngasse 5 \\
     A-1090 Wien, Austria \\

\vspace{.2in}

\end{center}

\vspace{.7in}

\thispagestyle{empty} 

\begin{abstract}
The complete renormalization of the weak Lagrangian to chiral order $q^2$
in heavy baryon chiral perturbation theory is performed
using heat kernel techniques.
The results are compared with divergences appearing
in the calculation of Feynman graphs for the nonleptonic hyperon decay
$\Lambda \rightarrow p \pi^-$ and an estimate for the size of the counterterm
contributions to the s-wave amplitudes in nonleptonic hyperon decays is given.
\end{abstract}

\vfill

\section{Introduction}  
Weak decays of hyperons
have been examined using effective field theory methods
for more than three decades \cite{DGH}, but still a number of mysteries
exist. The matrix elements of nonleptonic hyperon decays, {e.g.}, 
can be described in terms of two amplitudes --- the parity-violating
s-wave and the parity-conserving p-wave.
Chiral perturbation theory provides a framework
whereby these amplitudes can be expanded in terms of small
four-momenta and the current masses $m_q$ of the light quarks, $q=u,d,s$.
At lowest order in this expansion 
the amplitudes are expressed in terms of two unknown coupling
constants, so-called low-energy constants (LECs). However, there is no 
consensus for the determination of these two weak parameters. 
If one employs values which provide a good fit for the s-waves, one obtains
a poor description of the p-waves. On the other hand, a good p-wave 
representation yields a poor s-wave fit.
In order to overcome this problem, one must go beyond leading order.
In Refs. \cite{BSW,J}, a first attempt was
made in calculating the leading chiral corrections to such decays.
But the inability to fit s- and p-waves simultaneously remains
even after including the lowest nonanalytic contributions.
In a recent paper \cite{BH} a calculation was performed which included
{\it all} terms at one-loop order. An exact fit to the data was
possible but not unique, and other model-dependent assumptions had 
to be made in order to estimate the LECs.
Another intriguing possibility was examined by Le Yaouanc et al., who asserted
that a reasonable fit for both s- and p-waves can be provided by appending pole
contributions from $SU(6)\, (70,1^-)$ states to the s-waves \cite{ley}. Their
calculations were performed in a simple quark model and appear to be able to
provide a resolution of the s- and p-wave dilemma. The validity of this
approach has been confirmed within the framework of chiral perturbation theory,
but only after contributions from the lowest lying 1/2$^+$ baryon 
octet resonant states were also taken into account \cite{BH1}.

Another topic of interest are the radiative hyperon 
decays: $\Sigma^+\rightarrow
p\gamma,\Lambda\rightarrow n\gamma,$  etc.
Here, the primary problem has been
and remains to understand the size of the asymmetry parameter in
polarized $\Sigma^+\rightarrow p\gamma$ decay: $\alpha_\gamma=
-0.76\pm 0.08\label{eq:aa}$ \cite{pdb}.
The difficulty here is associated with the restrictions posed by Hara's
theorem, which requires the vanishing of this asymmetry in the U-spin
symmetric limit \cite{hara}.
Of course, in the real world, U-spin is broken and one should not be
surprised to find a nonzero value for the asymmetry --- what {\it is}
difficult to understand is its size.
Recent work involving the calculation of 
chiral loops has also not lead to a resolution, although
slightly larger asymmetries can be accomodated \cite{N}.
Within that work it was claimed, that 
in order to obtain a better understanding for the decays one should
account for {\it all} terms at one-loop order, 
i.e., including all counterterms, not just the leading
log corrections.
For a different approach including explicitly baryon resonances
which leads to improved agreement with experimental data from 
radiative hyperon decays see Refs. \cite{gav, BH2}. 
This might indicate that weak
hyperon decays cannot be described appropriately without the inclusion of
baryon resonances. We will not elaborate on this possiblity in the present
investigation. Rather, one of the main purposes of our paper is to estimate the
size of higher order counterterms which might shed some light in the
understanding of the origin of the discrepancy between theory and experiment
when baryon resonances are not taken into account.

At one-loop order, divergences appear
and are absorbed by infinite LECs from counterterms of the
same chiral order. The renormalized low-energy constants are
scale-dependent and measurable, i.e., they can in principle
be determined from a fit to some observables.
They satisfy renormalization group equations under scale changes
and, therefore, the choice of another scale leads to modified
values of the renormalized LECs. The divergences of the
generating functional determine the renormalization group equations
and the behavior of the renormalized LECs under scale changes.
The sum of the irreducible one-loop functional and the counterterm
functional is, of course, finite and scale-independent.
Some of these divergences
were treated in Ref. \cite{BH}, which, to our knowledge,
is the only work in the weak baryonic sector that performed 
renormalization explicitly.
Obviously, only a subset of the leading divergences were treated
in that work.
It is our aim to work out {\it all} leading divergences in the
generating functional of the weak baryon-meson Lagrangian,
thus extending the work of M{\"u}ller and Mei{\ss}ner \cite{guido}
in the strong sector, using two different techniques and to discuss
a few applications.
The complete divergence structure might be used as a check
in future calculations.

In the next section we present the weak Lagrangian at lowest order.
The generating functional to one-loop order is worked out in Sec. 3.
The divergent parts of the irreducible tadpole, self-energy
and the so-called eye-graph are isolated by using
heat kernel techniques. 
Sec. 4 contains a sample calculation of the divergent pieces
of the diagrams for the
nonleptonic hyperon decay $\Lambda \rightarrow p \pi^-$. 
The divergent pieces of these diagrams
are compared with the results from the heat kernel technique.
In Sec. 5 we give an estimate for some counterterm contributions to 
the s-wave amplitudes in nonleptonic hyperon decays and
we summarize our findings in Sec.6.
In the Appendix we extend the recently proposed
super heat kernel technique
to the weak effective Lagrangian.

\section{Effective Lagrangian}
We perform our calculations using the lowest order
effective Lagrangian within the heavy baryon formalism.
To this end, one writes down the most general relativistic
Lagrangian which is invariant under chiral and $CPS$ transformations.
Imposing invariance of the Lagrangian under the transformation $S$,
which interchanges down and strange quarks in the Lagrangian,
one can further reduce the number of counterterms.
We will work in the $CP$ limit so that all LECs are real.
This Lagrangian is then reduced to the heavy fermion limit
by the use of path integral methods, which deliver
the relativistic corrections 
as $1/ \!\! \mnod$ terms in higher orders. 
The baryons are described by a
four-velocity $v_{\mu}$ and a consistent chiral counting scheme emerges,
i.e., a one-to-one correspondence between the Goldstone boson loops
and the expansion in small momenta and quark masses.
However, we will not present the relativistic Lagrangian explicitly here, 
but rather quote only the form of the heavy baryon limit.

The pseudoscalar Goldstone fields ($\phi = \pi, K, \eta$) are collected in
the  $3 \times 3$ unimodular, unitary matrix $U(x)$, 
\begin{equation}
 U(\phi) = u^2 (\phi) = \exp \lbrace 2 i \phi / F_0 \rbrace
\end{equation}
with $F_0$ being the pseudoscalar decay constant (in the chiral limit), and
\begin{eqnarray}
 \phi =  \frac{1}{\sqrt{2}}  \left(
\matrix { {1\over \sqrt 2} \pi^0 + {1 \over \sqrt 6} \eta
&\pi^+ &K^+ \nonumber \\
\pi^-
        & -{1\over \sqrt 2} \pi^0 + {1 \over \sqrt 6} \eta & K^0
        \nonumber \\
K^-
        &  \bar{K^0}&- {2 \over \sqrt 6} \eta  \nonumber \\} 
\!\!\!\!\!\!\!\!\!\!\!\!\!\!\! \right) \, \, \, \, \, . 
\end{eqnarray}
Under SU(3)$_L \times$SU(3)$_R$, $U(x)$ transforms as $U \to U' =
LUR^\dagger$, with $L,R \in$ SU(3)$_{L,R}$.
The matrix $B$ denotes the baryon octet, 
\begin{eqnarray}
B  =  \left(
\matrix  { {1\over \sqrt 2} \Sigma^0 + {1 \over \sqrt 6} \Lambda
&\Sigma^+ &  p \nonumber \\
\Sigma^-
    & -{1\over \sqrt 2} \Sigma^0 + {1 \over \sqrt 6} \Lambda & n
    \nonumber \\
\Xi^-
        &       \Xi^0 &- {2 \over \sqrt 6} \Lambda \nonumber \\} 
\!\!\!\!\!\!\!\!\!\!\!\!\!\!\!\!\! \right)  \, \, \, ,
\end{eqnarray}
which under $SU(3)_L \times SU(3)_R$ transforms as any matter field,
\begin{equation} 
B \to B' = K \, B \,  K^\dagger
 \, \, \, ,
\end{equation}
with $K(U,L,R)$ the compensator field representing an element of the
conserved subgroup SU(3)$_V$.
At the order we are working, the effective Lagrangian has the form
\beq
{\cal L}_{\mbox{eff}}  =  \: {\cal L}_{\phi B} \: + \: 
        {\cal L}_{\phi B}^W  \:    + \: {\cal L}_{\phi} ,
   \eeq
where $ {\cal L}_{\phi} $
is the usual (strong and electromagnetic) mesonic 
Lagrangian, see,  e.g., Ref. \cite{GL1}.

For the strong meson-baryon Lagrangian ${\cal L}_{\phi B}  $
one writes
\beqa
{\cal L}_{\phi B} = {\cal L}_{\phi B}^{(1)}  
& = &
 \ci \, < \bar{B} [ v \cdot \nabla , B] > +            
D \, < \bar{B} S_{\mu} \{ u^{\mu}, B\} > 
+ F \, < \bar{B} S_{\mu} [ u^{\mu}, B] > 
\eeqa
with $2 S_\mu = \ci \gamma_5 \sigma_{\mu \nu} v^{\nu}$ 
denoting the Pauli-Lubanski spin vector, $<...>$ is defined as  the trace in flavor 
space and
the superscript denotes the chiral order.
At lowest order the 
meson-baryon Lagrangian contains two axial-vector couplings,
denoted by $D$ and $F$.
The covariant derivative $\nabla_\mu$ of the baryons is decomposed as
\beq
[\nabla_\mu,B] = \partial_\mu B + [\Gamma_\mu,B]
\eeq
with $\Gamma_\mu$ being the so-called chiral connection
\beq
\Gamma_\mu = \frac{1}{2} \Big[ u^\dagger( \partial_\mu - i r_\mu ) u + 
       u ( \partial_\mu - i l_\mu ) u^\dagger \Big] .
\eeq
The external fields $v_\mu , \, a_\mu$ appear in the combinations
$r_\mu = v_\mu + a_\mu$ and $l_\mu = v_\mu - a_\mu$.
The meson fields are summarized in the quantity 
\beq
u_\mu = \frac{1}{2} \Big[ u^\dagger( \partial_\mu - i r_\mu ) u - 
       u ( \partial_\mu - i l_\mu ) u^\dagger \Big] .
\eeq
We will also need the expression 
$\chi_+ = 4 B_0 {\cal M} + {\cal O}(\phi^2)$
with ${\cal M} = \mbox{diag}(m_u,m_d,m_s)$ being the quark mass matrix and 
$B_0 = - \langle 0 |\bar{q} q |0 \rangle/F_0^2$ the order parameter of the
spontaneous symmetry violation.

Having dealt with its strong counterpart,
the weak meson-baryon Lagrangian ${\cal L}_{\phi B}^W  $ at lowest order is
\beq
{\cal L}_{\phi B}^{W \, (0)}  =  \:
d \, < \bar{B}  \{ h_+ , B\} > + \:
f \, < \bar{B}  [ h_+ , B ] > .
\eeq
Here, we have defined
\beq
h_+ = u^\dagger h u + u^\dagger h^\dagger u ,
\eeq
with $h^a_b = \delta^a_2 \delta^3_b$ being the weak transition matrix.
Note that $h_+$ transforms as a matter field.

\section{Renormalization of the weak one-loop generating functional using 
standard heat kernel techniques}

In this section, we turn to the calculation of the complete one-loop generating functional in SU(3) heavy baryon chiral perturbation theory, 
i.e., to order $q^3$ for the strong interaction and 
to order $q^2$ for the weak sector in the small momentum expansion.
This extends the renormalization of the strong interacting functional
 \cite{guido} to the weak interaction. The method used in this section was first proposed by 
Ecker \cite{ecker} in the two flavor case.  We will focus our calculation
on applications that are of physical interest, so we neglect weak 
hyperon decays into two or more pions. 
Radiative and nonleptonic hyperon decays are of specific interest.

For the calculation it is useful to write the fields in terms
of the physical basis, e.g.,  $ B = B^a \lambda^a $
with $ < { \lambda^a }^\dagger \, \lambda^b > = \delta^{ab} $.
We can then write the meson-baryon interaction in the form
\beq
S_{\phi B} = \int d^4 x \bar{B}^a ( A^{ab}_{{\rm str}} +
 A^{ab}_{W} ) B^b
\eeq
with the strong and weak interaction pieces
$A^{ab}_{{\rm str}}$ and $A^{ab}_{W}$, respectively.
Following Refs. \cite{guido} and \cite{ecker} 
one has to expand
\beq
{\cal L}_\phi^{(2)} +  {\cal L}_\phi^{(4)} - \bar{R}^a  \,
[A^{ab}_{(1), {\rm str}}  +  A^{ab}_{(0), W } ]^{-1} \, R^b
\label{Zexp}
\eeq
in the functional integral around the classical
solution, $u_{\rm cl} = u_{\rm cl}[j]$, which is obtained by the variation
$\delta \, \int d^4x \,{\cal L}_{\phi}^{(2)} / \delta U$ at lowest order.  
Here, $j$ collectively denotes the external fields $v_\mu, a_\mu$ and the quark
mass matrix ${\cal M}$.
The baryon source fields have also been decomposed into a light component,
denoted by $R$,  and a heavy 
component where the heavy components are 
not needed for a consistent renormalization and will therefore be omitted.
One thus arrives at a set of irreducible and reducible diagrams.
From Eq. (\ref{Zexp}) we immediately derive
\beq
{\cal L}_\phi^{(2)} +  {\cal L}_\phi^{(4)} 
- \bar{R}^a  \, [A^{ab}_{(1), {\rm str}}  ]^{-1} \, R^b 
\, + \, \bar{R}^a  \, [A^{ab}_{(1), {\rm str}}  ]^{-1} \,[ A^{ab}_{(0), W }  
 ] \,  [A^{ab}_{(1), {\rm str}}  ]^{-1} ]  \, R^b \, \, ,
\label{Zexpfin}
\eeq
where the first baryon term leads to the well known tadpole, self-energy 
contributions and the last baryon term is the new eye-graph part with a 
weak insertion on the intermediate baryon line. This problem was solved
in Refs. \cite{guido2} and \cite{thesis} for the strong interaction with one 
insertion from the second order Lagrangian.   
As in the mesonic sector, we
choose the fluctuation variables $\xi$ in a symmetric form \cite{GL1},
\beq
\xi_R = u_{\rm cl} \, \exp\{i \, \xi /2\} \,\,\, , \quad 
\xi_L = u_{\rm cl}^\dagger \, \exp\{-i \, \xi /2\} \,\,\, , 
\eeq
with $\xi^\dagger = \xi$ traceless 3$\times$3 matrices. Consequently,
we also have
\beq
U =  u_{\rm cl} \, \exp\{i \, \xi \} \, u_{\rm cl} \,\,\, .
\eeq  
At second order in $\xi$, the covariant derivative $\nabla_\mu$, 
the chiral connection $\Gamma_\mu$, the axial-vector $u_\mu$
and $h_+$ take the form
\beqa
\Gamma_\mu &=& \Gamma_\mu^{\rm cl} + \frac{1}{4} \,[\, u_\mu^{\rm cl}, \xi
\, ] + \frac{1}{8} \, \xi 
\stackrel{\leftrightarrow}{\nabla}_\mu^{\rm cl}\, \xi + {\cal
  O}(\xi^3) \,\,\, ,
\nonumber \\ 
{} [ \nabla_\mu^{\rm cl} , \xi ] &=& \partial_\mu \, \xi + [ \, \Gamma_\mu^{\rm
  cl}, \xi \, ] \,\, ,\,\, \xi \stackrel{\leftrightarrow}{\nabla}_\mu^{\rm
  cl} \xi = \xi 
[ \nabla_\mu^{\rm cl} , \xi ] - [ \nabla_\mu^{\rm cl} , \xi ] \xi
\,\,\,\, ,
\nonumber \\
u_\mu &=& u_\mu^{\rm cl} - [ \nabla_\mu^{\rm cl} , \xi ] + \frac{1}{8} \,
[\, \xi, [\, u_\mu^{\rm cl}, \xi\,]\,] + {\cal O}(\xi^3) \nonumber \\ 
h_+ &=&  h_+^{\rm cl} + \frac{i}{2} [ h_+^{\rm cl}, \xi ]  +
\frac{1}{8} [ \xi , [ h_+^{\rm cl} , \xi ]] + + {\cal O}(\xi^3)  \,\,\, .
\label{nablaclsu3}
\eeqa
Inserting this into the expression for $A^{ab}$ and retaining
only the terms up to and including order $\xi^2$ gives
\beqa
A_{(1), {\rm str} }^{ab} &=& A_{(1), {\rm str}}^{ab, \, \rm cl}
+\frac{i}{4}\,<{\lambda^a}^\dagger \, [\,[ v \cdot u_{\rm cl}, \xi
\,], \lambda^b\,]> - D/F< {\lambda^a}^\dagger \, ( [S \cdot \nabla_{\rm
  cl} , \xi ] , \lambda^b \, )_\pm > \nonumber  \\
&+& \frac{i}{8}\,<{\lambda^a}^\dagger \, [\,\xi \, v \cdot 
\stackrel{\leftrightarrow}{\nabla}_{\rm cl} \xi ,
\, \lambda^b\,]> + \frac{1}{8} D/F< {\lambda^a}^\dagger \, ( 
\,[\,\xi, [\,S \cdot u_{\rm cl}, \xi \, ]\,], \lambda^b )_\pm >
+ {\cal O}(\xi^3) ,  \nonumber \\ 
A_{(0),W}^{ab} &=& A_{(1), W }^{ab, \, \rm cl}
+\frac{i}{2}\,<{\lambda^a}^\dagger \, [\, h_+^{\rm cl}, \xi
\,], \lambda^b\,]> +  \frac{1}{8} d/f< {\lambda^a}^\dagger \, ( 
\,[\,\xi, [\,  h_+^{\rm cl}   , \xi \, ]\,], \lambda^b )_\pm >
+ {\cal O}(\xi^3) \,\, , 
\label{Aab1}
\eeqa
where we have introduced the compact notation
\beq
D/F ( \lambda^a, \lambda^b)_\pm = D \{\lambda^a, \lambda^b\} + 
F[\lambda^a, \lambda^b] .
\eeq
The corresponding generating functional reads 
\beqa
Z_{\rm irr}[j,\bar{R}^a, R^e] &=& \int d^4x \, d^4x' \, d^4y \, d^4y' \, 
\bar{R}^a (x) \, S_{(1), {\rm str}}^{ac, \,\rm cl} (x,y) \nonumber\\
& & \times \bigl[ \, \Sigma_{{\rm tad} }^{cd}(y,y') \,  \delta(y-y')
+ \Sigma_{{\rm self}}^{cd} (y,y')+ \Sigma_{{\rm eye}}^{cd} (y,y') \,
 \bigr] \, S_{(1), {\rm  str}}^{de, \, \rm cl}(y',x')
\, R^e(x') \nonumber\\
\label{Zirrsu3}
\eeqa
in terms of the tadpole, self-energy and eye-graph functionals.
Here, $S^{\rm cl}_{(1), {\rm str}}$ is the full 
classical fermion propagator with the
weak interactions turned off.
The functionals  are given by
\beqa
\label{S12Vi}
\Sigma_{{\rm self}}^{ab} &=& -\frac{2}{F_0^2} \, V_i^{ac} \, G_{ij} \, 
[A_{(1), {\rm str}}^{cd,cl}]^{-1} \, V_j^{db} 
=  -\frac{2}{F_0^2} \, V_i^{ac} \, G_{ij} \, 
S_{(1), {\rm str}}^{cd,cl} \, V_j^{db} \nonumber \\
\Sigma_{{\rm eye}}^{ab} &=& \frac{2}{F_0^2} \, V_i^{ac} \, G_{ij} \, 
[A_{(1), {\rm str}}^{ce,cl}] \, [A_{(0),W}^{ef,cl}] \, 
[A_{(1), {\rm str}}^{fd,cl}]^{-1} \, V_j^{db} 
 \nonumber \\
\Sigma_{{\rm tad}}^{ab}
 &=& \frac{1}{8F_0^2} \biggl\{ D/F < {\lambda^a}^\dagger \, (
 \, [ \, \lambda^i_G \, , \, [ \, S \cdot u^{cl} , \lambda^j_G \,] \, ]\,  , 
\lambda^b \,)_\pm > \, G_{ij} \nonumber \\
&& \qq
 +  d/f < {\lambda^a}^\dagger \, (
 \, [ \, \lambda^i_G \, , \, [ \, h_+ , \lambda^j_G \,] \, ]\,  , 
\lambda^b \,)_\pm > \, G_{ij} \nonumber \\
& & \qq + i < {\lambda^a}^\dagger \, [ \,
\lambda^i_G (G_{ij} v \cdot \stackrel{\leftarrow}{d}_{jk} 
- v \cdot d_{ij} G_{jk} ) \lambda^k_G ,
\lambda^b \, ] > \biggr\} 
\eeqa
with the following definitions of the vertices
\beqa 
V_i^{ab} & = & V_{i, {\rm str}}^{ab}  + V_{i, W}^{ab}  \,\,\,\,\, , \,\,\,
V_{i, {\rm str}}^{ab}  =   V_{i, {\rm str}}^{ab(1)} + 
V_{i, {\rm str}}^{ab(2)} \nonumber \\
V_{i, {\rm str}}^{ab(1)} & = & \frac{i}{4\sqrt{2}} < {\lambda^a}^\dagger \, [\, [ v
\cdot u^{cl} , \lambda^i_G \,],  \, \lambda^b \, ] > \,\,\,\, , \,\,\,
V_{i, {\rm str}}^{ab(2)}  =  - \frac{D/F}{\sqrt{2}}
< {\lambda^a}^\dagger \, ( \, \lambda^j_G \, S \cdot d_{ji}, \lambda^b
\, )_\pm > \nonumber  \\
V_{i , W}^{ab} & = & \frac{d/f}{2\sqrt{2}}
< {\lambda^a}^\dagger \, ( \, [  h_+^{\rm cl} , \lambda^j_G ] \, , \lambda^b
\, )_\pm > 
\eeqa
where $i,j,k,a,b,c= 1, \ldots ,8$
and $\lambda^i_G$ denote Gell-Mann's SU(3)
matrices, which are related to the ones in the physical basis by
$\lambda_p = (\lambda^4_G+i \lambda^5_G)/2$, $\lambda_n = (\lambda^6_G+
i \lambda^7_G)/2$, etc.  
The quantity $G_{ij}$ is the full meson propagator \cite{GL1}
\beq
G_{ij} =  ( d_\mu \, d^\mu \, \delta^{ij} + \sigma^{ij} \, )^{-1}
\label{Mespropsu3}
\eeq
with 
\beqa
[\nabla^\mu_{\rm cl}\, , \xi] &=& \frac{1}{\sqrt{2}} \lambda_G^j \,
d^\mu_{jk} \, \xi_k \,\,  \, , 
\quad \xi = \frac{1}{\sqrt{2}} \, \lambda_G^i \, \xi_i \,\, , \nonumber \\
d^\mu_{ij} &=& \delta_{ij} \, \partial^\mu + \gamma^\mu_{ij} \,\, ,
\stackrel{\leftarrow}{d}^\mu_{ij} = \delta_{ij}\,
\stackrel{\leftarrow}{\partial}^\mu - \gamma_{ij}^\mu \,\, , 
\nonumber \\
\gamma^\mu_{ij} &=& -\frac{1}{2} < \Gamma^\mu_{\rm cl} \, [ \, \lambda_G^i,
\lambda_G^j \, ] > \,\,\, , \nonumber \\
\sigma^{ij} &=& \frac{1}{8} <[ \, u^{\rm cl}_\mu , \lambda_G^i \, ][ \,
\lambda_G^j  , u_{\rm cl}^\mu   \, ] + \chi_+ \, \{ \, \lambda_G^i ,
\lambda_G^j \,\} > \,\,\, .
\label{MesHKsu3}
\eeqa
Note that the differential operator $d_{ij}$ is related to the
covariant derivative $\nabla^\mu_{\rm cl}$ and it acts on the meson
propagator $G_{ij}$. 
The connection $\gamma_\mu$ defines a field strength tensor,
\beqa
\gamma_{\mu \nu} &=& \partial_\nu \, \gamma_\mu - \partial_\mu \,
\gamma_\nu + [ \gamma_\mu , \gamma_\nu ] \,\,\, , 
\label{mesfssu3}
\eeqa
where we have omitted the flavor indices. 
We are now in a position to derive 
the divergences of the one-loop generating functional 
for weak and strong interaction after contracting trace indices 
in flavor space.
In the heat kernel representation, the divergences appear as simple poles
in $ \epsilon= 4-d $.
The beta functions of the strong
interacting functional are given in Ref. \cite{guido} and the new divergent 
contribution of the weak interaction 
 can be cast in the form (rotated back to Minkowski space) 
\beq
\Sigma_{{\rm weak}}^{ab , {\rm div}} (y,y) = \frac{1}{(4 \pi F_0 )^2}  
\frac{1}{\epsilon} \, \Biggl[ \, \hat{\Sigma}_{{\rm tad}}^{ab} (y,y) 
+   \, \, \hat{\Sigma}_{{\rm self}}^{ab} (y,y) 
+ \, \, \hat{\Sigma}_{{\rm eye}}^{ab} (y,y) \Biggl] \,\,\, ,
\label{divergent}
\eeq
where  $\hat{\Sigma}^{ab} (y,y)$  are finite monomials in the
fields of chiral dimension two.
Let us start with the tadpole contribution which is given by
\beqa
\hat{\Sigma}_{{\rm tad}}^{ab} (y,y) &=& -\frac{1}{4}\, d/f \, \biggl\{
-\frac{3}{2} < {\lambda^a}^\dagger  \, 
( \{ h_+ ,  \chi_+ \} , \lambda^b)_\pm > \nonumber \\
&-&  < {\lambda^a}^\dagger  \, ( h_+ , \lambda^b )_\pm><  \chi_+> \biggl\} \,  
 - \, \frac{1}{2}d \, < \, {\lambda^a}^\dagger \,\lambda^b \, > < h_+ \, \chi_+ >   \, \,\,\, . 
\label{tad} 
\eeqa 
Notice that we have neglected terms of order $u^2$ which contain at least two 
external pions,
since they do not contribute to nonleptonic and radiative hyperon decays. 
The self-energy contribution is given by
\beqa
\hat{\Sigma}_{{\rm self}}^{ab} (y,y) &=& 6 (Df+Fd)  < {\lambda^a}^\dagger  \,
 \{  [ h_+ , S^\mu \Gamma_{\mu\nu} v^\nu ] ,  \lambda^b \}  >  
\nonumber \\
&+&  (\frac{10}{3} Dd + 6 Ff)  < {\lambda^a}^\dagger  \,
 [ [ h_+ , S^\mu \Gamma_{\mu\nu} v^\nu ] ,  \lambda^b ]  > 
\nonumber \\
 &+&  \frac{3}{2} f < {\lambda^a}^\dagger  \{ v \cdot u , \{ h_+ , [i v \cdot 
 \nabla ,  \lambda^b] \} \}  >  + \mbox{ h.c. }  
\nonumber \\
&+& \frac{3}{2} f < {\lambda^a}^\dagger  [ v \cdot u , [ h_+ , [i  v \cdot 
 \nabla ,  \lambda^b] ]]  >  + \mbox{ h.c. }  
\nonumber \\ 
  &+&  \frac{3}{2} d < {\lambda^a}^\dagger  \{ v \cdot u , [ h_+ , [i  v \cdot 
 \nabla ,  \lambda^b] ] \}  >  + \mbox{ h.c. }  
\nonumber \\
&+& \frac{3}{2} d < {\lambda^a}^\dagger  [ v \cdot u , \{ h_+ , [ i v \cdot 
 \nabla ,  \lambda^b] \} ]  >  + \mbox{ h.c. }  
\nonumber \\
&+& 3d  < {\lambda^a}^\dagger \{ [ h_+ , [i v \cdot \nabla , v \cdot u ]] 
\,  , \, \lambda^b \}  > 
+ 3d  < {\lambda^a}^\dagger \{ [ [ i v \cdot \nabla , h_+ ]  , v \cdot u ] 
\,  , \, \lambda^b \}  >  \nonumber \\ 
&+& 3f  < {\lambda^a}^\dagger [ [ h_+ , [i v \cdot \nabla , v \cdot u ]] 
\,  , \, \lambda^b]   > 
+ 3f  < {\lambda^a}^\dagger [ [ [i v \cdot \nabla , h_+ ]  , v \cdot u ] 
\,  , \, \lambda^b]  >    \nonumber \\
&+& 4f < {\lambda^a}^\dagger [i v \cdot \nabla , \lambda^b] > < v \cdot u \, 
 h_+ >  \, +   \mbox{ h.c. }  \nonumber  \\
&+& 4f < {\lambda^a}^\dagger  h_+ > 
< v \cdot u  \, [i v \cdot \nabla , \lambda^b] >   \, +   \mbox{ h.c. } \, .
\eeqa 
Here, $\Gamma_{\mu\nu}$ denotes the field strength tensor of the fields
$\Gamma_{\mu}$.
The last contribution stems from the eye-graph and takes the form
\beqa \label{eye}
\hat{\Sigma}_{{\rm eye}}^{ab} (y,y) &=& [S^\mu , S^\nu ] \biggl\{ 
 ( -4 (D^2+F^2) f  - 8DFd )  < {\lambda^a}^\dagger  \, \lambda^b >
  <   h_+ \, \Gamma_{\mu\nu} >  \nonumber \\
 &+& 4f (D^2-F^2)  <   {\lambda^a}^\dagger  \,  h_+ > < \, \Gamma_{\mu\nu} \, 
  \lambda^b >  + \mbox{ h.c. }  \nonumber \\ 
&+& (3DFd - \frac{3}{2} f (D^2+F^2) ) 
 <   {\lambda^a}^\dagger  \, \{  \Gamma_{\mu\nu} \, \{ h_+ , \, 
  \lambda^b \} \} >  + \mbox{ h.c. }  \nonumber \\ 
&+& ( -\frac{1}{3} DFd + \frac{3}{2} f (D^2+F^2) ) 
 <   {\lambda^a}^\dagger  \, [  \Gamma_{\mu\nu} \, [ h_+ , \, 
  \lambda^b ]] >  + \mbox{ h.c. }  \nonumber \\ 
&+& (3DFf - \frac{3}{2} d (D^2+F^2) ) 
 <   {\lambda^a}^\dagger  \, \{  \Gamma_{\mu\nu} \, [ h_+ , \, 
  \lambda^b ] \} >  + \mbox{ h.c. }  \nonumber \\ 
&+& ( -\frac{17}{3} DFf + \frac{3}{2} d (D^2+F^2) ) 
 <   {\lambda^a}^\dagger  \, [  \Gamma_{\mu\nu} \, \{ h_+ , \, 
  \lambda^b \} ] >  + \mbox{ h.c. }    \,\, \biggl\} \nonumber \\
&+& ( 6 D^2d - 18 F^2d + 36 DFf  )
 < [{\lambda^a}^\dagger , v \cdot \stackrel{\leftarrow}\nabla ] 
\{ h_+ , [ v \cdot \nabla ,\lambda^b] \} > \nonumber \\
&+& ( -10 D^2f - 18 F^2f + 20 DFd  )
 < [{\lambda^a}^\dagger , v \cdot \stackrel{\leftarrow}\nabla ] 
 [ h_+ , [ v \cdot \nabla ,\lambda^b] ] > \nonumber \\
&+& ( -2 D^2d  +6 F^2d -12 DFf  )
 < {\lambda^a}^\dagger \, \{ [ v \cdot \nabla , \, [ 
v \cdot \nabla ,  h_+  ]] , \, \lambda^b \} > \nonumber \\
&+& ( \frac{10}{3} D^2f  +6 F^2f - \frac{20}{3}  DFd  )
 < {\lambda^a}^\dagger \, [ [ v \cdot \nabla , \, [ 
v \cdot \nabla ,  h_+  ]] , \, \lambda^b ] >  \nonumber \\
&-& \frac{3}{16}  \biggl\{   ( \frac{68}{9} D^2d + 4F^2d -8DFf) 
< {\lambda^a}^\dagger \, \{ h_+ ,  \lambda^b \} > < \chi_+ >  \nonumber \\
&+&  ( \frac{68}{9} D^2f + 4F^2f -8DFd) 
< {\lambda^a}^\dagger \, [ h_+ ,  \lambda^b ] > < \chi_+ >  \nonumber \\
&+&  (  8 (D^2 + F^2)d + 8DFf) 
< {\lambda^a}^\dagger \,  \lambda^b  > < h_+ \, \chi_+ >  \nonumber \\
&+&  (  \frac{136}{9} D^2d - 8 F^2d ) 
< {\lambda^a}^\dagger \, h_+  > <\chi_+ \,  \lambda^b  >  + \mbox{ h.c. } 
 \nonumber \\
&+&  ( - \frac{23}{3} D^2d - 6 DFf  + 3 F^2d ) 
< {\lambda^a}^\dagger \, \{ \chi_+ \, \{  h_+ , \,  \lambda^b \} \}  > 
 + \mbox{ h.c. }  \nonumber \\
&+&  (  -3 D^2d  + \frac{2}{3} DFf  - 3 F^2d ) 
< {\lambda^a}^\dagger \, [ \chi_+ \, [  h_+ , \,  \lambda^b ]] > 
 + \mbox{ h.c. }  \nonumber \\
&+&  (  - \frac{7}{3} D^2f  - \frac{2}{3} DFd  + 3 F^2f ) 
< {\lambda^a}^\dagger \, \{ \chi_+ \, [  h_+ , \,  \lambda^b ] \}  > 
 + \mbox{ h.c. }  \nonumber \\
&+&  (  -3 D^2f  + \frac{2}{3} DFd  - 3 F^2f ) 
< {\lambda^a}^\dagger \, [ \chi_+ \, \{  h_+ , \,  \lambda^b \} ] > 
 + \mbox{ h.c. } \biggl\}
\nonumber \\
&+&  (  - \frac{68}{9} D^3d -12 D^2Ff  -12 DF^2d -4 F^3f ) 
< {\lambda^a}^\dagger \,  [ i v \cdot \nabla ,  \lambda^b ] > 
 < S \cdot u \, h_+ >   + \mbox{ h.c. }  \nonumber \\
&+&  (   -4 D^3d +4 D^2Ff  +4 DF^2d -4 F^3f ) 
< {\lambda^a}^\dagger  \, h_+ >  < S \cdot u \, 
  [ i v \cdot \nabla ,  \lambda^b ] >  + \mbox{ h.c. } \nonumber \\
&+&  (  3 D^3d +3  D^2Ff  +3 DF^2d - 3 F^3f ) 
< {\lambda^a}^\dagger  \, \{  h_+  , \, \{ S \cdot u \, 
  [ i v \cdot \nabla ,  \lambda^b ] \} \}  >  + \mbox{ h.c. } \nonumber \\
&+&  (  \frac{1}{3} D^3d +  5 D^2Ff  + \frac{13}{3} DF^2d - 3 F^3f ) 
< {\lambda^a}^\dagger  \, [  h_+  , \, [ S \cdot u \, 
  [ i v \cdot \nabla ,  \lambda^b ] ] ]  >  + \mbox{ h.c. } \nonumber \\
&+&  (  -\frac{7}{3} D^3f +   D^2Fd  +3 DF^2f - 3 F^3d ) 
< {\lambda^a}^\dagger  \, \{  h_+  , \, [ S \cdot u \, 
  [ i v \cdot \nabla ,  \lambda^b ] ] \}  >  + \mbox{ h.c. } \nonumber \\
&+&  (  - D^3f  - \frac{1}{3} D^2Fd  + 3 DF^2f - 3 F^3d ) 
< {\lambda^a}^\dagger  \, [  h_+  , \, \{ S \cdot u \, 
  [ i v \cdot \nabla ,  \lambda^b ] \} ]  >  + \mbox{ h.c. }
 \nonumber \\ 
&+&  (  \frac{2}{9} D^3f  - \frac{26}{9} D^2Fd  -2 DF^2f + 2 F^3d ) 
< {\lambda^a}^\dagger  \, \{ [[ i v \cdot \nabla  ,  h_+ ]  , \,  S
\cdot u ] \, ,\lambda^b \}  >   \nonumber \\ 
&+&  (  \frac{2}{3} D^3d  - 2 D^2Ff  + \frac{14}{9} DF^2d + 2 F^3f ) 
< {\lambda^a}^\dagger  \, [ [[ i v \cdot \nabla  ,  h_+ ]  , \,  S
\cdot u ] \, ,\lambda^b ]  >   \nonumber \\ 
&+&  (  -\frac{2}{9} D^3f  + \frac{26}{9} D^2Fd  +2 DF^2f -2 F^3d ) 
< {\lambda^a}^\dagger  \, \{ [  h_+   , \, [ i v \cdot \nabla  , S
\cdot u ]] \, ,\lambda^b \}  >   \nonumber \\ 
&+&  ( - \frac{2}{3} D^3d  + 2 D^2Ff  + \frac{22}{9} DF^2d - 2 F^3f ) 
< {\lambda^a}^\dagger  \, [ [ h_+   , \, [ i v \cdot \nabla  ,  S
\cdot u ]] \, ,\lambda^b ]  > 
\,\,\, .
\eeqa

\noindent 
It is now straightforward to pin down
the full weak counterterm
Lagrangian at order $q^2$. We also use the curvature relation
for $\Gamma_{\mu \nu}$ 
\beq
\Gamma_{\mu \nu} = \frac{1}{4}[u_\mu, u_\nu]  - \frac{i}{2} F^+_{\mu
  \nu} 
\eeq
where $F_{\mu \nu}^+ = u F_{\mu \nu}^L u^\dagger + 
u^\dagger F_{\mu \nu}^R u$ with
$F_{\mu \nu}^{L/R}$ being the field strength tensors related to the external
fields $l_\mu$ and $r_\mu$, respectively.
The generating functional can be renormalized by introducing the
counterterm Lagrangian for the strong interaction \cite{guido}
\beq \label{LCTsu3}
{\cal L}_{\phi B}^{(3)\, } (x) = \frac{1}{(4 \pi F_0 )^2} \, \sum_i
d_i \, \bar{B}^{a} (x) \, \tilde{O}^{ab}_{i, {\rm str}}(x) \, B^{b} (x)
\eeq
where the $d_i$ are dimensionless coupling constants and the field
monomials $\tilde{O}^{ab}_{i, {\rm str}}  (x)$ are of order $q^3$ 
and by introducing for the weak interaction the counterterm Lagrangian 
\beq \label{LCTweak}
{\cal L}_{\phi B}^{(2)\, W} (x) = \frac{1}{(4 \pi F_0 )^2} \, \sum_i
h_i \, \bar{B}^{a} (x) \, \tilde{O}^{ab}_{i, W} (x) \, B^{b} (x)
\eeq 
where the $h_i$ are dimensionless coupling constants and the field
monomials $\tilde{O}^{ab}_{i, W} (x)$ are of order $q^2$.  
The low-energy 
constants  are decomposed into
\beqa \label{29}
d_i &=& d_i^r (\mu ) + (4 \pi)^2 \, \beta_i \, L(\mu ) \,\,\, \nonumber \\
h_i &=& h_i^r (\mu ) + (4 \pi)^2 \, \beta_i \, L(\mu ) \,\,\, ,
\eeqa
with $\mu$ being the scale introduced in dimensional regularization and
\beq
L(\mu ) = \frac{\mu^{d-4}}{(4 \pi)^2} \Big( \frac{1}{d-4} - \frac{1}{2} [ \log
(4 \pi) +1 - \gamma] \Big) ,
\eeq 
where $\gamma = 0.5772215...$ is the Euler-Mascheroni constant.
The $\beta_i$ are dimensionless functions of $F$, $D$ and $f$, $d$ constructed
such that they cancel the divergences of the one-loop functional. 
The renormalized LECs $d_i^r (\mu) \, \, ( h_i^r (\mu) )$
 are measurable quantities. They
satisfy the renormalization group equations
and therefore, the choice of another scale leads to modified
values of the renormalized LECs.
We remark that the scale-dependence in the counterterm Lagrangian is,
of course, balanced by the scale-dependence of the renormalized
finite one-loop functional for observable quantities.

The next two sections are devoted to the application of the formulae 
derived in
the present investigation. First, we will show how the divergences can be used
to check results obtained from the ordinary computation of Feynman diagrams. 
Second, the scale-dependence of the renormalized LECs as given in Eq. 
(\ref{29}) can be employed to make an estimate of the size of such higher order
counterterms.

\section{A sample calculation}
In this section we will discuss the renormalization of the
nonleptonic hyperon decay $\Lambda \rightarrow p \pi^-$ by
calculating explicitly the Feynman graphs and comparing them with
the results obtained in the previous section.
In the rest frame of the heavy baryon, $ v_{\mu} = (1,0,0,0) $,
the decay amplitude reduces to the non-relativistic form
\beqa
{\cal A}( B_i \rightarrow B_j \, \pi) & = &
\bar{u}_{B_j} \, \Big\{ \, {\cal A}_{ij}^{(s)} + \, 
S \cdot k \, {\cal A}_{ij}^{(p)}\,  \Big\}u_{B_i} \qq ,
\eeqa
where $k$ is the outgoing momentum of the pion.
Here, $ {\cal A}_{ij}^{(s)}$ is the parity-violating s-wave amplitude and
$ {\cal A}_{ij}^{(p)}$ is  the corresponding parity-conserving p-wave term.
In this frame, the energy of the outgoing pion is
\beq
v \cdot k = \frac{1}{2 m_i} \Big( m_i^2 - m_j^2 + M_\pi^2 \Big)
\eeq
and the energy of the decaying hyperon in the heavy baryon formulation
can be written as
\beq
v \cdot p =  m_i - \mnod .
\eeq
Here, $m_{i,j}$ are the physical masses of the baryons and $\mnod$
is the mass of the baryon octet in the chiral limit.
Since the baryon masses are analytic to linear order in the quark masses
we see that $ v \cdot p$ and $ v \cdot k$ count effectively as order
${\cal O}(q^2)$.
We will therefore restrict ourselves to the computation of the
one-loop divergences to this decay which are proportional to the quark
mass matrix and do not contain the momenta of the external particles.
In our example we present only the renormalization of the weak
vertices.
The divergent structure of the purely strong baryonic Lagrangian which also
contributes to this decay is already given in Ref. \cite{guido}.

We start by renormalizing the p-wave amplitude. There are four
Feynman graphs contributing to the p-wave and containing mass-dependent
divergences (see Fig. 1). For the renormalization
procedure it is sufficient to consider only the irreducible tadpole
from Figures 1a and 1b and the irreducible eye-graph from Figs. 1c and 1d
by neglecting the parts from the internal baryon propagator and the
strong vertex. The contributions $P^a$ and $P^b$ 
of the irreducible tadpole in Figures 1a and 1b to the decay then read 
\beqa
P^a & = & - \frac{i}{4 F_0^2} ( d - f)
         L(\mu) \: \Big( 3 M_\pi^2 + 6 M_K^2 + 3 M_\eta^2 \Big) \no \\
P^b & = &  \frac{i}{4 F_0^2} \frac{1}{\sqrt{6}} ( d + 3 f)
         L(\mu) \: \Big( 3 M_\pi^2 + 6 M_K^2 + 3 M_\eta^2 \Big) .
\eeqa
The divergent pieces can be recovered by using the
results from the previous section. Employing the counterterms
from Eq. (\ref{tad}) and
using the Gell-Mann-Okubo relation for the pseudoscalar mesons,
$ 4 M_K^2 = 3 M_\eta^2 + M_\pi^2 $, which is consistent
to the order we are working, cancels the divergences
from the calculation of the one-loop graphs. Note that the last term in
Eq. (\ref{tad}) does not contribute here, since $ \langle
 h_+ \chi_+\rangle  =0 $.

The mass-dependent divergences from diagrams 1c and 1d read
after neglecting the internal baryon propagator and the strong vertex 
\beqa
P^c & = & \frac{i}{F_0^2} L(\mu) \bigg( M_\pi^2 \Big[ -\frac{2}{3} D^2 d
      - 8 DFf + 6 F^2 d - \frac{10}{3} D^2 f + 4 DFd -6 F^2 f \Big] \no \\
 & &  \qq + M_K^2 \Big[ -\frac{7}{3} D^2 d
  - 10 DFf + 3 F^2 d - \frac{5}{3} D^2 f + 6 DFd -3 F^2 f \Big] \bigg)\no \\
P^d & = & \frac{i\sqrt{6}}{F_0^2} L(\mu)
          \bigg( M_\pi^2 \Big[ \frac{14}{9} D^2 d
      - 2 DFf - \frac{4}{3} D^2 f + \frac{4}{3} DFd  \Big] \no \\
 & &  \qq \: + M_K^2 \Big[ -\frac{19}{18} D^2 d + 5 DFf - \frac{3}{2} F^2 d 
  -  \frac{7}{6} D^2 f + \frac{11}{3} DFd - \frac{9}{2} F^2 f \Big] \bigg) .
\eeqa
The divergent pieces are in agreement with the mass-dependent
divergences in Eq. (\ref{eye}).

The renormalization of the s-wave is somewhat more subtle.
In this case it turns out that in addition to the simple irreducible
tadpole diagram in Fig. 2a contributing to the s-wave there are
three more diagrams (2b,c,d) which lead to the same divergent structure.
The pertinent divergent contributions which are proportional to the
quark mass matrix for Fig. 2a read
\beq
S^a =  \frac{1}{24 \sqrt{3}  F_0^3} \: ( d + 3f )
          L(\mu) \: \Big( 5 M_\pi^2 + 16 M_K^2 + 9 M_\eta^2 \Big) 
\eeq
and for Figures 2c and 2d 
\beq
S^c + S^d = - \frac{1}{4 \sqrt{3} F_0^3} \: ( d + 3f )
     L(\mu) \: \Big( 2 M_\pi^2 + M_K^2  \Big) .
\eeq
The sum of graphs 2a, 2c and 2d leads to a divergence
structure different than that obtained from the tadpole occuring in
the p-wave. Agreement between
s- and p-waves is only achieved if one includes the 
$Z$-factor of the pion which is divergent and, therefore, contributes to
the divergences for nonleptonic hyperon decay. This was not treated 
correctly in Ref. \cite{BH}. 
From an explicit calculation of the pion $Z$-factor including
the mesonic Lagrangian of fourth chiral order ${\cal L}_\phi^{(4)}$
one obtains
\beq
\sqrt{Z_\pi} = - \frac{1}{F_0^2} L(\mu) \frac{2}{3} \Big( 2 M_\pi^2 +
             M_K^2 \Big) \: + \: \mbox{finite term} 
\eeq
which is multiplied by the pertinent tree level contribution
to the decays
\beq
- \frac{1}{2 \sqrt{3} F_0} ( d + 3f) .
\eeq
Expanding $h_+$ in Eq. (\ref{tad}) 
in terms of the meson fields
\beq
h_+ = h - \frac{i}{F_0} [\phi,h] + {\cal O}(\phi^2, h^\dagger)
\eeq
the term which is linear in the meson fields reproduces the divergences
of the one-loop contributions to the s-wave discussed above. 
Note that the pseudoscalar decay constant in the chiral limit
$F_0$ is finite and does not lead to additional divergences.

Finally, Fig. 2e leads to the divergences
\beqa
S^e & = & \frac{\sqrt{3}}{F_0^3} L(\mu)
         \bigg( M_\pi^2 \Big[ \frac{14}{9} D^2 d
      - 2 DFf - \frac{4}{3} D^2 f + \frac{4}{3} DFd  \Big] \no \\
 & &  \qq \qq + M_K^2 \Big[ -\frac{19}{18} D^2 d + 5 DFf - \frac{3}{2} F^2 d 
  -  \frac{7}{6} D^2 f + \frac{11}{3} DFd - \frac{9}{2} F^2 f \Big] \bigg) .
\eeqa
This result agrees with the mass dependent divergences in Eq. (\ref{eye}).

\section{Importance of counterterms}
Finally, we would like to address the issue of the importance of higher order
counterterms for nonleptonic hyperon decays. 
There exist seven such transitions: 
$\Sigma^+  \rightarrow n \, \pi^+ \, , \, 
\Sigma^+  \rightarrow p \, \pi^0 \, , \, 
\Sigma^- \rightarrow n \, \pi^- \, , \, 
\Lambda \rightarrow p \, \pi^- \, ,  \,
\Lambda \rightarrow n \, \pi^0 \, , \,
\Xi^- \rightarrow  \Lambda \, \pi^-\, , \, \mbox{and} \, 
\Xi^0 \rightarrow  \Lambda \, \pi^0 $.
Isospin symmetry of the strong interactions implies the relations 
\beqa \label{iso}
&&
{\cal A}(\Lambda \rightarrow p \, \pi^-)
+ \sqrt{2} \, {\cal A}(\Lambda \rightarrow n \, \pi^0) = 0 \no \\
&&
{\cal A}(\Xi^- \rightarrow  \Lambda \, \pi^-)
+ \sqrt{2} \, {\cal A}(\Xi^0 \rightarrow  \Lambda \, \pi^0) = 0 \no \\
&&
\sqrt{2} \, {\cal A}(\Sigma^+ \rightarrow p \, \pi^0)
+ {\cal A}(\Sigma^- \rightarrow n \, \pi^-)
- {\cal A}(\Sigma^+ \rightarrow n \, \pi^+) = 0
\eeqa
which hold both for s- and p-waves.
We choose
$\Sigma^+  \rightarrow n \, \pi^+ \, , \, 
\Sigma^- \rightarrow n \, \pi^- \, , \, 
\Lambda \rightarrow p \, \pi^- \, , \, \mbox{and} \,
\Xi^- \rightarrow  \Lambda \, \pi^- $
to be the four independent decay amplitudes which are not related by isospin.
As mentioned in the introduction, both a good s- and p-wave fit cannot be
achieved just by taking the lowest order couplings $d, f$ and chiral
corrections into account. This suggests that one has to consider higher order
counterterms, but there are so many of them that they cannot be determined
uniquely \cite{BH}.
However, as we will show in this section 
the divergent structure of the one-loop functional can be 
used to give
an estimate of the size of these counterterm contributions.
In order to keep the presentation more compact, we restrict ourselves to 
the case of the s-waves.
Neglecting divergences which are proportional to $v \cdot k$ and $v \cdot q$
(see last section) the possible counterterms are linear in $\chi_+$ and read
\beqa
{\cal L}_{\phi B}^{W \, (2,br)}  
& = &
h_1 \bigg\{ \tr \Big( \bar{B}  \{ h_+ , \{  \chi_+, B\} \} \Big)  
         +\tr \Big( \bar{B}  \{ \chi_+, \{ h_+ , B \} \}  \Big) \bigg\} \no \\ 
& + & 
h_2 \bigg\{ \tr \Big( \bar{B}  [ h_+ , [ \chi_+, B] ] \Big)  
         +\tr \Big( \bar{B}  [ \chi_+, [ h_+ , B ]]  \Big) \bigg\} \no \\ 
& + & 
h_3 \bigg\{ \tr \Big( \bar{B}  [ h_+ , \{ \chi_+, B\} ] \Big)  
         +\tr \Big( \bar{B}  \{ \chi_+, [ h_+ , B ]\}  \Big) \bigg\} \no \\ 
& + & 
h_4 \bigg\{ \tr \Big( \bar{B}  \{ h_+ , [ \chi_+, B] \} \Big)  
         +\tr \Big( \bar{B}  [ \chi_+, \{ h_+ , B \} ]  \Big) \bigg\} \no \\ 
& + & 
h_5 \bigg\{ \tr \Big( \bar{B} h_+ \Big)  \tr \Big( \chi_+  B \Big)
+ \tr \Big(\bar{B} \chi_+ \Big) \tr \Big( h_+ B \Big)
\bigg\} \no \\ 
& + & 
h_{6} \tr \Big( \bar{B}  [ h_+ , B ] \Big) \tr \Big( \chi_+ \Big) +
h_{7} \tr \Big( \bar{B}  \{ h_+ , B ]\} \Big) \tr \Big( \chi_+ \Big) .
\eeqa
Note that we did not make any use of Cayley-Hamilton identities in order to
have the same set of counterterms as in Eqs. (\ref{tad}) and (\ref{eye}).
After renormalization of the mass-dependent divergences the contributions of
these counterterms read 
\beqa
{\cal A}^{(s)}_{\Sigma^+ n }  &=& 0 \nonumber \\
{\cal A}^{(s)}_{\Sigma^- n}  &=& 
 \frac{2\sqrt{2}}{F_\pi} \Big(  M_\pi^2  [ h_1^r - h_2^r -h_3^r + h_4^r 
 - \frac{1}{2}  h_6^r + \frac{1}{2} h_7^r ]  \nonumber \\
&& \qq + M_K^2 [ h_1^r + h_2^r -h_3^r - h_4^r - h_6^r + h_7^r ] \Big) \no \\
{\cal A}^{(s)}_{\Lambda p}  &=& 
 \frac{2}{\sqrt{3}F_\pi} \Big(  M_\pi^2  [ 3 h_1^r - 3 h_2^r + h_3^r - h_4^r 
      + 2 h_5^r - \frac{3}{2} h_6^r - \frac{1}{2} h_7^r ]  \nonumber \\
&& \qq + M_K^2 [- 5 h_1^r + 3 h_2^r -7 h_3^r + h_4^r 
      - 2 h_5^r - 3 h_6^r -  h_7^r ] \Big) \nonumber \\
{\cal A}^{(s)}_{\Xi^- \Lambda } &=& 
 \frac{2}{\sqrt{3}F_\pi} \Big(  M_\pi^2  [ 3 h_1^r - 3 h_2^r - h_3^r + h_4^r 
      + 2 h_5^r + \frac{3}{2} h_6^r - \frac{1}{2} h_7^r ]  \nonumber \\
&& \qq + M_K^2 [- 5 h_1^r + 3 h_2^r + 7 h_3^r - h_4^r 
      - 2 h_5^r + 3 h_6^r -  h_7^r ] \Big) 
\eeqa
where we have not shown explicitly the scale-dependence of the $h_i^r$.
At the order we are working, we replace $F_0$ by the pion decay constant
$F_\pi = 92.4$ MeV since the differences between the two show up at higher
orders only.
The dependence on a chosen scale $\mu_1$ 
of these counterterm contributions is, of course,
compensated by the scale-dependence of the renormalized finite one-loop
functional for observable quantities.
The choice of another scale $\mu_2$ leads to modified values of the
renormalized LECs according to the equation
\beq
h_i^r (\mu_2 ) =  h_i^r (\mu_1 ) + \beta_i \, \log \frac{\mu_1}{\mu_2} .
\eeq
Varying the scale from the $\rho$-meson mass $M_\rho = 770$ MeV to the $\Delta$
mass $M_\Delta = 1232$ MeV one obtains the following differences in the
counterterm contributions for the s-waves, in units of 10$^{-7}$,
\beqa
  {\cal A}^{(s)}_{\Sigma^- n}(\mu_\rho) - 
       {\cal A}^{(s)}_{\Sigma^- n}(\mu_\Delta) &=&  -0.77  \qq (4.27)  \no \\
 {\cal A}^{(s)}_{\Lambda p}(\mu_\rho) - 
       {\cal A}^{(s)}_{\Lambda p}(\mu_\Delta) &=&  0.60  \qq (3.25)  \no \\
{\cal A}^{(s)}_{\Xi^- \Lambda }(\mu_\rho) - 
       {\cal A}^{(s)}_{\Xi^- \Lambda }(\mu_\Delta) &=&  1.94  \qq (-4.51) 
\eeqa
which is about 19\% for $\Sigma^- \rightarrow n \, \pi^- \, , \, 
\Lambda \rightarrow p
\, \pi^- $ and even 43\% for $\Xi^- \rightarrow  \Lambda \, \pi^-$;
the experimental values for the decays being given in brackets.
There are no contributions at one-loop order
for the decay $\Sigma^+ \rightarrow n \, \pi^+$ \cite{BH}, 
neither from the counterterms nor from the 
chiral loops;
we have therefore omitted its presentation here.
Assuming that the differences of these counterterm contributions for different
realistic choices of the scale give an estimate of their absolute value
$\delta {\cal A}^{(s)}$, we obtain,  in units of 10$^{-7}$,
\beqa
| \delta {\cal A}^{(s)}_{\Sigma^- n } | &=& 0.77 \no \\
| \delta {\cal A}^{(s)}_{\Lambda p } | &=& 0.60 \no \\
| \delta {\cal A}^{(s)}_{\Xi^-  \Lambda }| &=& 1.94 .
\eeqa
While the counterterm contributions to the s-waves seem to be well behaved for 
$\Sigma^- \rightarrow n \, \pi^-$ and $\Lambda \rightarrow p \, \pi^- $,
our calculation indicates that they might be significant for 
$\Xi^- \rightarrow  \Lambda \, \pi^-$.

\section{Summary}
In this paper, we have performed the chiral invariant renormalization
of the weak effective baryon-meson
Lagrangian up to one-loop order within heavy baryon chiral perturbation
theory. The complete set of counterterms at leading one-loop order $q^2$ with
$q$ being an external momentum or meson mass has been constructed.
This extends work by M{\"u}ller and Mei{\ss}ner \cite{guido}, who considered
the strong $SU(3)$ baryon-meson Lagrangian. The present calculation 
has been performed both using 
the standard heat kernel formalism and the recently proposed 
super heat kernel method which has the advantage of simplifying
the calculation in intermediate steps. We also 
compared our results with a direct calculation of the Feynman graphs
for the nonleptonic hyperon decay $\Lambda \rightarrow p \pi^-$.
It turns out that the divergences contained in
the $Z$-factor of the pion are essential for achieving agreement
between the renormalization of s- and p-wave amplitudes.
Since, to our knowledge, there exists only one calculation in the weak
baryon-meson sector where some of the divergences at order ${\cal O}(q^2)$ 
have been evaluated
\cite{BH}, our work might serve as a reference to check further calculations
in the future.

The low-energy constants of higher order counterterms contain, in general,
divergent pieces which cancel the divergences from one-loop graphs. The finite
remainder of the counterterms is scale-dependent and compensates the 
scale-dependence of the one-loop functional to give scale independent 
expressions for
physical quantities. For radiative and nonleptonic hyperon decays there exist
more counterterms than there are experimental data \cite{BH, N}, so that the
counterterms cannot be determined uniquely from experiment.
Nevertheless, the scale-dependence of the finite remainder of the LECs after
renormalization can be used to give an estimate of the size of the counterterm
contributions. Since the divergent structure of the one-loop functional
determines the scale-dependence of the renormalized LECs, we are able to give
such estimates and as an example we have chosen the contributions of the
counterterms linear in the quark mass matrix ${\cal M}$ to the s-waves of the
independent nonleptonic hyperon decays that are not related by isospin.
We find that for the s-waves
the contributions from these counterterms relatively to the
experimental values
range from 19\% for $\Sigma^- \rightarrow n \, \pi^-$ and 
$\Lambda \rightarrow p\, \pi^- $
up to 43\% for the decay $\Xi^- \rightarrow \Lambda \pi^-$.

\section*{Acknowledgments}
Useful discussions with H. Neufeld and S. Steininger are gratefully
acknowledged. This work was supported in part by the Deutsche
Forschungsgemeinschaft and the BMBF.

\section*{Appendix: Renormalization using the super heat kernel formalism}

We provide an alternative renormalization prescription by 
using the super heat kernel formalism as proposed by Ecker, Gasser, and
Neufeld \cite{EGN}. 
This formalism has already been applied to the effective two- and three-flavor
Lagrangian in Refs.
\cite{Neu} and \cite{MN}, respectively, where it served as a check for previous
work \cite{guido,ecker}.
We use this approach to calculate the divergences of
the weak effective Lagrangian. This method  allows us to verify
our previous calculation presented in Sec.3.

The fluctuation action generated by the
lowest order meson-baryon Lagrangian has the general form
\beq
\label{HBfluc}
S^{(2)} = -\dfrac{1}{2}\xi^T(D_{\mu}D^{\mu} + Y)\xi +
  \ol \eta ( \alpha + \beta_{\mu} D^{\mu}) \xi  
 + \xi^T (\ol \delta -\ol \beta_{\mu} \D^{\mu}) \eta 
+ \ol \eta i v^{\mu} \D_{\mu} \eta  ~,
\eeq
where $\psi (\eta)$ are the bosonic (fermionic) fluctuations and 
\beq
\ba{ll}
D_\mu = \del_\mu + X_\mu~, & \D_\mu = \del_\mu + f_\mu~, \\[4pt]
\ol \delta = \ol \alpha - \ol {\D_\mu \beta^\mu}~,     \\
v^2 = 1~, & v \cdot \beta = 0~. \ea
\label{explic}
\eeq
$X_\mu$, $Y$, and $f_\mu$ are bosonic (matrix) fields, 
whereas $\alpha$ and $\beta_\mu$ are fermionic objects. The form of $\ol
\delta$ in Eq. (\ref{explic}) is required by the reality of Eq. (\ref{HBfluc}).
Apart from the condition $v \cdot \beta = 0 $
no further assumption about the terms entering in Eq. (\ref{HBfluc}) is
made. 
To be specific, in HBCHPT we find:
\noindent
\beqa
X_\mu &=& \gamma_\mu + g_\mu \, , \,\, 
(g_\mu)^{ij}  = 
-i \frac{v_\mu}{8 F_0^2} < \bar B \, [ [\lambda^i_G , \lambda^j_G ],  B ] >
\, , \,\,\, i, j=1,...,8 \, , \nonumber \\
Y &=& \sigma  + s \, , \,\,\,  \nonumber \\
s^{ij} &=& -\frac{D/F}{4F_0^2} < \bar B \ ( [ \lambda^i_G , [ S \cdot u, \lambda^j_G]], B )_\pm >  
 \,\, 
 -\frac{d/f}{4F_0^2} < \bar B \ ( [ \lambda^i_G , [ S \cdot u , \lambda^j_G ]], B )_\pm > \, ,  \nonumber \\
f_\mu &=& f_\mu^{{\rm str}}  + f_\mu^{W}  \, , \nonumber \\
f_\mu^{ab, {\rm str}} &=&  < {\lambda^a}^\dagger [ \Gamma_\mu , \lambda^b ] > 
 -i v_\mu D/F < {\lambda^a}^\dagger ( S \cdot u  , \lambda^b )_\pm >   \, ,
\nonumber \\
f_\mu^{ab, {\rm W}} &=& -i v_\mu  d/f < {\lambda^a}^\dagger ( h_+ , \lambda^b )_\pm > 
\,\,  ,  a=1,...,8  \, , \nonumber \\
\alpha^{ai} &=& \frac{i}{4F_0}  < {\lambda^a}^\dagger [ [ v \cdot u  , 
\lambda^i_G ] , B ]  >  + \frac{i}{2F_0}  < {\lambda^a}^\dagger [ [ h_+  , 
\lambda^i_G ] , B ]  >   \, , \nonumber \\
(\beta_\mu)^{ai}  &=&  - \frac{D/F}{F_0}  \, S_\mu \, 
 < {\lambda^a}^\dagger ( \lambda^i_G  , B )_\pm  >  
\eeqa
with the definitions of Section 3. \\
Details of super heat kernel calculation in SU(3) can be found in Ref.
\cite{MN}. We only use the final result to calculate  the 
structure of the divergences. Using these definitions we easily
derive from the following formula
the pertinent beta functions and the counterterms 
\beqa
W_{L=1}^{\rm div}|_{\ol \Gamma \ldots \Gamma} &=& 
\dfrac{i}{48 {\pi}^2 (d-4)} \int d^4x \mbox{ tr }
\left \{ - 12 \, \ol \alpha \, v \cdot {\wh \nabla} \alpha 
 + 6 \left [ \ol \alpha \beta_\mu X^{\mu \nu} v_\nu
         +\ol \beta_\mu \alpha X^{\mu \nu} v_\nu \right ] \right. \nn
&&\left. - 3 \left [ \ol \beta \cdot \beta \, v \cdot {\wh \nabla} Y
         + 2 \, \ol \beta_\mu (v \cdot {\wh \nabla} \beta^{\mu}) Y \right
] 
- 4 \ol \beta_\mu (v \cdot {\wh \nabla})^3 \beta^{\mu}
+ \ol \beta \cdot \beta \, {\wh \nabla}_\mu X^{\mu \nu}
v_\nu \right. \nn
&&\left.  + 6 \, \ol \beta_\mu (v \cdot {\wh \nabla} \beta_\nu)
X^{\mu \nu}
   + 4 \, \ol \beta _\mu \beta_\nu \, v \cdot {\wh \nabla} X^{\mu
\nu}
   + 2 \, \ol \beta_\mu \beta_\nu {\wh \nabla}^{\mu} X^{\nu \rho}
      v_{\rho} \right \}~. 
\label{result}
\eeqa 
where  $X^{\mu \nu}$ is the field strength tensor of the fields $ X^\mu$.
This result can be used since the weak interaction
has a relatively simple structure and does not contain any divergences
which act on the bosonic or fermionic fluctuation variables. The new
eye-graph contribution can be evaluated by redefining the covariant
derivative in the most economic way, i.e., $f_\mu = f_\mu^{{\rm str}} 
 + f_\mu^{W}  $. This shows that the
eye-graph without any derivative 
is immediately obtained in the super heat kernel formalism as it is 
the case in the standard heat kernel method. 
But the situation
changes dramatically in the presence of derivatives.
So far the result is given in a very compact form, and both the beta functions
and the corresponding monomials must be evaluated by contracting the
various trace indices. In the SU(3) case this is the most tedious part of the
calculation and therefore, we skip the presentation of these
technicalities. After a lot of cumbersome 
algebra one ends up with the 
result presented in the last section.

\newpage


\section*{Figure captions}

\begin{enumerate}

\item[Fig.1] Given are the diagrams which contribute
         to the mass-dependent divergences of the p-wave
         in the nonleptonic hyperon decay $\Lambda \rightarrow p \pi^-$. 
         Solid and dashed lines denote baryons and
         pseudoscalar mesons, respectively. The solid square represents
         a weak vertex and the solid circle denotes a strong vertex.

\item[Fig.2] Given are the diagrams which contribute
         to the mass-dependent divergences of the s-wave
         in the nonleptonic hyperon decay $\Lambda \rightarrow p \pi^-$. 
         Solid and dashed lines denote baryons and
         pseudoscalar mesons, respectively. The solid square represents
         a weak vertex and the solid circle denotes a strong vertex.

\end{enumerate}

\newpage

\begin{center}
 
\begin{figure}[bth]
\centering
\centerline{
\epsfbox{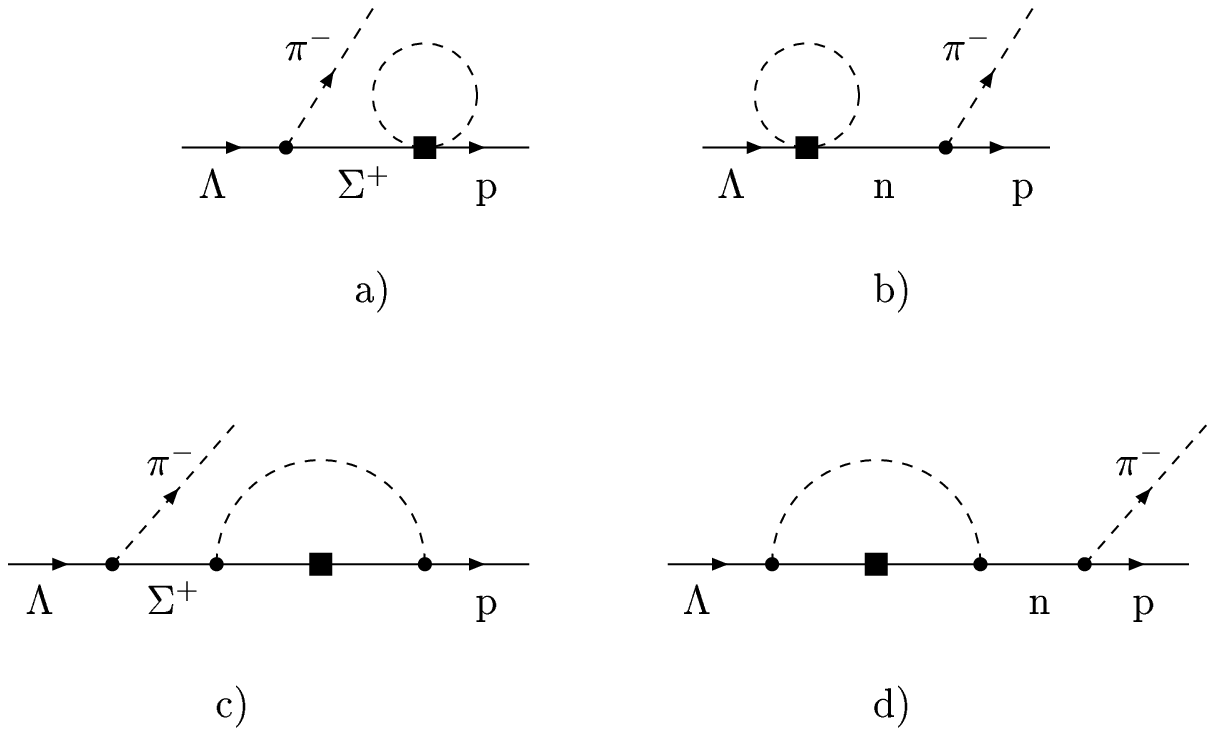}}
\end{figure}

\vskip 0.7cm

Figure 1

\newpage

\begin{figure}[tbh]
\centering
\centerline{
\epsfbox{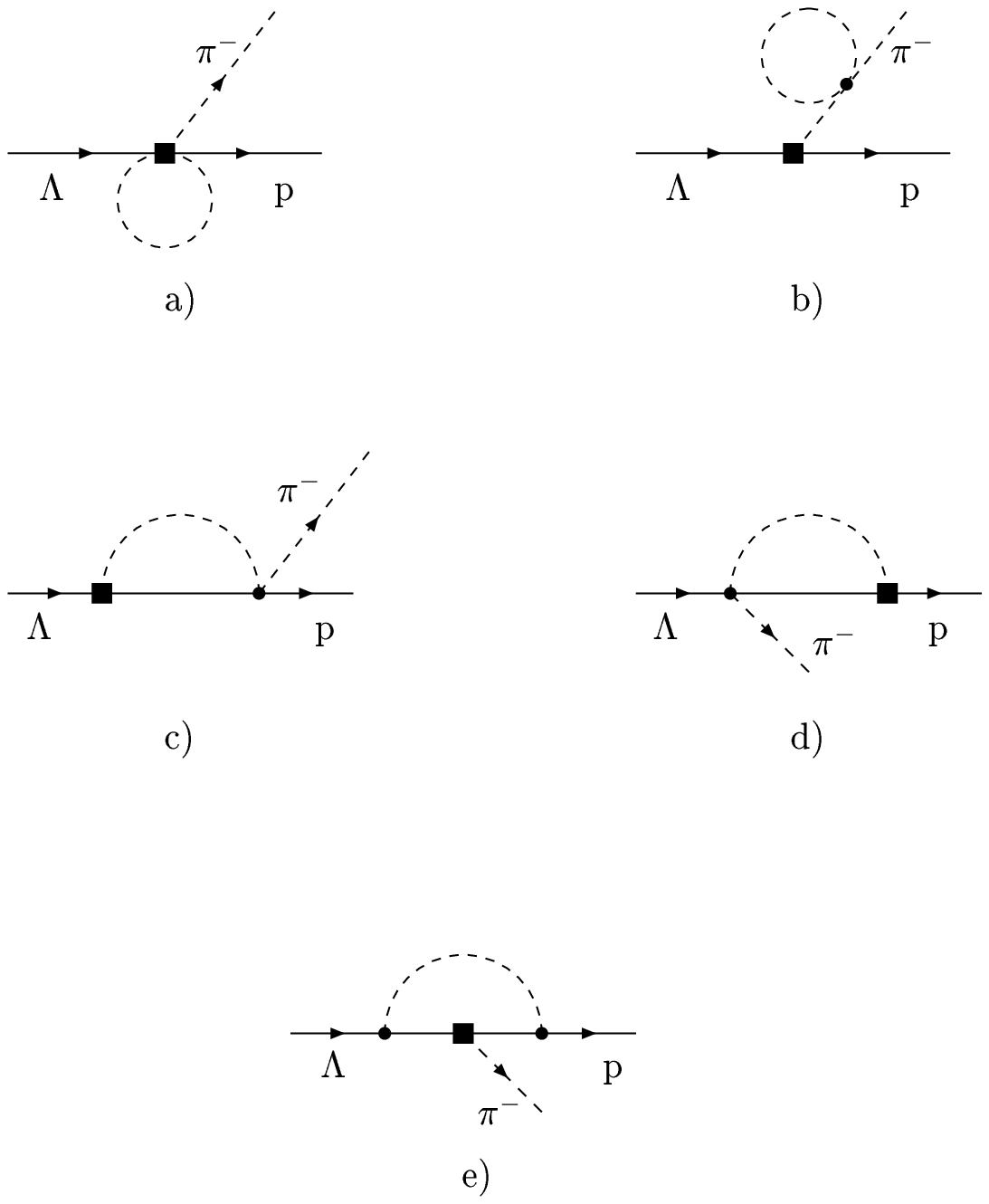}}
\end{figure}

\vskip 0.7cm

Figure 2

\end{center}

\end{document}